# Are Four-Fermion Operators Relevant for the Fermion Mass Problem?

B. Holdom

*Department of Physics*
*University of Toronto*
*Toronto, Ontario*
*Canada, M5S 1A7*

holdom@utcc.utoronto.ca

Four-fermion operators may have large anomalous scaling and become relevant operators in some strongly interacting gauge theories. We present a detailed model which illustrates some of the implications of such operators for the generation of quark and lepton masses. Such operators would originate at high scales where quarks and leptons experience a new strong interaction, but no unbroken technicolor interaction is required. The breakdown of both the new gauge symmetry and electroweak gauge symmetry is associated with a dynamical TeV mass for fourth family quarks. Among the new physics signatures are anomalous (chromo)magnetic moments and their flavor-changing counterparts.

# 1 Introduction

The existence of elementary scalar fields allows the values of the quark and lepton masses to be encoded into the values of the Yukawa couplings, which in turn can originate at very high energies. The alternative, dynamical symmetry breaking, brings the physics of fermion mass generation down to relatively low scales. Despite the infrastructure needed to have scalar fields at low scales (the introduction of supersymmetry and some mechanism to break supersymmetry), the idea that fermion masses originate at some near Planckian energy remains the more popular notion. The prime reason for this seems to be the concept of gauge coupling constant unification and its approximate agreement with current data. In this paper we would like to consider another possible mechanism for transmitting the effects of very high energy physics into the fermion mass spectrum. Although our proposal requires strong interactions and thus an accompanying set of dynamical assumptions, the mechanism may be intrinsically more natural than the postulate of light scalar fields, and therefore easier to implement in an economical framework.

When the standard model is regarded as a low energy effective theory of some Planckian theory, its renormalizability follows from the fact that only renormalizable operators remain relevant at low energies. If there are only weak couplings then the set of relevant terms is determined by naive power counting. But in strongly interacting theories the situation may be less trivial, and operators which are naively irrelevant may in fact be relevant. In particular we would like to explore the possibility that effects of physics at very high scales can feed down to low scales via 4-fermion operators having large anomalous scaling. We would like these operators to be such that they produce quark and lepton masses in the presence of the dynamical fermion masses responsible for electroweak symmetry breaking. We have in mind a new gauge symmetry which acts on normal quarks and leptons and which completely breaks down at some scale. Above this scale it strongly affects the scaling of at least some operators composed of quark and lepton fields. The emphasis of this paper will be to explore possible implications of such operators, especially in regard to the fermion mass problem, and not in trying to understand the origin of such operators in some strongly interacting sector at high energies.

Four-fermion operators in the role of relevant operators have been studied [1] in some detail in strongly interacting quenched QED in the ladder approximation. In that case it was shown that four-fermion operators had to be introduced in order to obtain a consistent, cut-off independent renormalization of the theory. There are some strong indications that this is more than just an artifact of the ladder approximation. An anomalous dimension $\gamma_m = 1$ for the mass operator $\bar{\psi}\psi$ (and thus an anomalous dimension at least close to 2 for a 4-fermion operator $\bar{\psi}\psi\bar{\psi}\psi$) seems to follow as soon as the coupling is large enough to produce chiral symmetry breaking. An argument for this based on the operator product expansion may be found in [2]. Another argument based on general properties the Schwinger-Dyson (SD) equation beyond the ladder approximation in quenched QED may be found in [3]. The essential requirement for a theory to have $\gamma_m = 1$ over any significant range of scales is a small or vanishing $\beta$-function, and thus a theory with a nontrivial infrared fixed-point may be of most interest. Although we must simply assume the existence of such a theory, we contrast the lack of fine tuning here to the case of a 'critically-strong' 4-fermion interaction in the absence of gauge interactions. In that case an effective low-energy scalar degree of freedom may be produced, at the cost of extreme fine-tuning of the strength of the 4-fermion interaction. It is in this sense that the postulate of relevant 4-fermion operators may be more natural than the postulate of light scalar fields.

We have mentioned that the strong interaction responsible for keeping certain 4-fermion operators relevant may be approaching a nontrivial infrared fixed-point. Since this interaction is not responsible for producing a small (TeV) mass (as for example in the walking technicolor case), the actual value of the coupling at this fixed-point does not have to be close to a critical value. As long as it is larger than some critical value then it should produce an anomalous dimension close to 2 for the 4-fermion operators. A related issue is how strong interactions can keep 4-fermion operators relevant over a large range of scales without these same interactions generating fermion mass. Of course if all fermions develop a mass at the high scale then there is no interesting low energy theory. We note that in a chiral gauge theory a dynamical reason exists for why some 4-fermion operators may be favored over mass; some 4-fermion operators may break no gauge symmetries while all masses do. Thus the gauge interactions in some chiral gauge theory may be sufficient to cause certain four-fermion operators to be relevant, but not sufficient to produce mass. For this to hold over a wide range of scales again seems to require a nontrivial infrared fixed point.

Returning to the issue of coupling constant unification, in dynamical symmetry breaking schemes it is difficult to imagine how any kind of coupling constant unification could survive, since the standard model gauge group usually becomes embedded into some larger group well before a typical unification scale. This seems to be necessary to avoid the problem of Goldstone bosons; the standard model gauge group is not large enough to either absorb or give mass to all the possible Goldstone bosons in otherwise realistic models. For example some kind of Pati-Salam unification is usually invoked close to the extended technicolor (ETC) scale to give sufficient mass to some neutral technipions.

This problem could be eliminated if there exist relevant 4-fermion operators which explicitly break global symmetries, thus removing unwanted Goldstone bosons. If this permits the standard model gauge group to survive up to some very high scale, one then wonders about the effect the new strong interactions will have on the running of the standard model gauge couplings. Here we note that as long as any new fermions come in standard model families, the **relative** running of the three standard model couplings is not affected at leading order in these couplings, to all orders in the new strong couplings. Thus the basic tendency for the three standard model couplings to become more equal at some large scale remains. Although this might provide some motivation for our proposal, we will not pursue this observation any further here. In fact some of the operators we discuss will be generated on scales just two or three orders of magnitude above the electroweak scale. Other operators including those responsible for eliminating global symmetries may be generated on much higher scales, but we leave open for now the question of exactly where they originate.

Perhaps the main problem facing theories of dynamical symmetry breaking is the issue of a large $t$ quark mass in association with a small $\delta\rho$ parameter. When a dynamical origin of quark mass is discussed it is usually assumed that the effective 4-fermion operator responsible takes the form $(\bar{H}_L H_R)(\bar{q}_R q_L) + h.c.$ (we will omit the '$h.c.$' from now on), where $H$ is some technifermion receiving a dynamical mass from technicolor interactions. In ETC theories the isospin breaking required in this operator to produce a large $t$ mass typically also shows up in the 4-technifermion operators $(\bar{H}_L H_R)(\bar{H}_R H_L)$, thus often implying that the technifermion sector will make a large contribution to $\delta\rho$. The problem is severe [4] for walking technicolor because in that case a ($\bar{H}_L H_R$)($\bar{H}_R H_L$) operator is enhanced more than a ($\bar{H}_L H_R$)($\bar{q}_R q_L$) operator. In fact in the walking limit it is believed [5] that the former is an example of an operator with an anomalous dimension close to 2, while that of the latter is close to 1. What we are looking for is the opposite situation, in which the operator contributing to the $t$ mass is enhanced by anomalous scaling at least as much or more than any other operator which contributes to $\delta\rho$.

The operator responsible for the $t$ mass must remain relevant to physics on scales as low as a TeV. This implies that at least some remnant of the new gauge symmetry must survive down to a similar scale. All fermions involved in the operator responsible for the $t$ mass must feel this strong gauge interaction. But eventually any such new gauge symmetry must be broken (otherwise the $b$ quark to which it would also couple would be stable). We are thus led to consider a new strong gauge interaction broken close to a TeV.

The model we shall present below will rely on 4-fermion operators of a form different than normally considered. That is, in place of operators of the form ($\bar{H}_L H_R$)($\bar{q}_R q_L$) we are led to consider operators of the form ($\bar{q}_L H_R$)$\epsilon$($\bar{H}_L q_R$), where again $H$ is a TeV mass fermion. The $SU(2)_L$ indices are contracted with the antisymmetric 2×2 matrix $\epsilon$, and such an operator can again feed mass from $H$ to $q$. These chirality-changing operators must have a nonperturbative origin in the same way that mass has a nonperturbative origin in dynamical symmetry breaking. Initial attempts to study the SD equation for dynamical 4-point functions of this *LRLR* form indicate that they tend to form for gauge couplings very similar to those needed to produce a dynamical mass [6]. Other chirality-changing 4-point functions, for example of the tensor-tensor form, seem to require larger couplings. We note that chiral symmetries may allow *LRLR* operators while not allowing mass, and this again points to the dynamics of chiral gauge theories.

In the next section we will identify the operators responsible for the $t$ and $b$ masses, and the gauge interaction responsible for enhancing these operators down to the TeV scale. We will also specify a larger gauge symmetry present above 100–1000 TeV, and additional operators which may be generated by this dynamics. We show how these latter operators cause mass mixing between the second and third families, and we will also argue that these operators induce the symmetry breakdown at a TeV. After describing this basic framework we describe some phenomenological consequences in section 3. These include the effects of $Z$ mixing with a new gauge boson having flavor dependent couplings, and additional anomalous gauge couplings induced by the operators discussed in section 2. In section 4 we continue to elaborate on various operators and show their relation to a realistic set of masses and mixings for four families. We conclude in section 5.

## 2 Basic Framework

It is helpful to rename our fields to explain why the $(\bar{q}_L H_R)\epsilon(\bar{H}_L q_R)$ operator is interesting. We will consider two quark doublets $Q \equiv (U, D)$ and $\underline{Q} \equiv (\underline{U}, \underline{D})$ which are not the mass eigenstates, but which will end up describing third and fourth family quarks. In terms of these fields our operator contains the following two $SU(2)_L \times U(1)_Y$ symmetric pieces.

$$(\bar{Q}_L D_R)\epsilon(\underline{\bar{Q}}_L \underline{U}_R) \quad (\mathcal{B})$$
$$(\bar{Q}_L U_R)\epsilon(\underline{\bar{Q}}_L \underline{D}_R) \quad (\tilde{\mathcal{B}}) \tag{1}$$

We shall require that $\bar{Q}_L Q_R$ and $\underline{\bar{Q}}_L \underline{Q}_R$ are gauge singlets with respect to the new strong gauge interaction, and for this reason we may expect an anomalous scaling enhancement of these operators. On the other hand the $\bar{H}_L H_R$ mass operator for the heavy fermion(s) is now written as $\underline{\bar{Q}}_L Q_R$, and this will **not** be invariant under the new gauge interaction. This means that there is a dynamical connection between the formation of the $\underline{\bar{Q}}_L Q_R$ mass and the breakdown of the new strong interaction. The $\underline{\bar{Q}}_L Q_R$ mass causes electroweak symmetry breaking and thus this mass must be of order a TeV. This is also an appropriate scale for the breakdown of the new strong interaction because we would like it to cause anomalous scaling of operators down to that scale. We will describe at the end of this section a possible origin for this gauge symmetry breaking.

The TeV mass quarks will correspond to the fourth family quarks $t'$ and $b'$. The $t$ and $b$ masses on the other hand correspond to $\bar{Q}_L \underline{Q}_R$, and thus the $\mathcal{B}$ and $\tilde{\mathcal{B}}$ operators produce the $t$ and $b$ masses in the presence of the fourth family $b'$ and $t'$ masses respectively. In order to keep the number of quark doublets to a minimum we will choose the new interaction to be a $U(1)$. We refer to this gauge boson as the $X$ boson. The $Q$ and $\underline{Q}$ must have equal and **opposite** vectorial $X$ charges, so that $\bar{Q}_L Q_R$ and $\underline{\bar{Q}}_L \underline{Q}_R$ are neutral while $\underline{\bar{Q}}_L Q_R$ and $\bar{Q}_L \underline{Q}_R$ are not. Fig. 1 illustrates the form of the diagrams which produce the anomalous scaling enhancement of the *t* and *b* masses.[1]

To see how the first two families could fit into this picture consider a larger gauge symmetry on a higher scale, into which the $U(1)_X$ becomes embedded. We take this gauge symmetry to be $U(1)_V \times SU(2)_V$ and we assume that it breaks to $U(1)_X$ at scale $\Lambda$ (which may be in the 100–1000 TeV range). We let the doublets $Q_i$ and $\underline{Q}_i$ with $SU(2)_V$ index $i$ transform as $(+, 2)$ and $(-, \bar{2})$ respectively under $U(1)_V \times SU(2)_V$. Our previous $Q$ and $\underline{Q}$ now correspond to $Q_1$ and $\underline{Q}_1$, and $U(1)_X$ corresponds to a combination of $U(1)_V$ and the $\sigma_3$ generator of $SU(2)_V$. The doublets $Q_2$ or $\underline{Q}_2$ have no $X$ charge and will describe quarks in the first two families. We will not speculate here as to the cause of the breakdown of $U(1)_V \times SU(2)_V$ to $U(1)_X$.

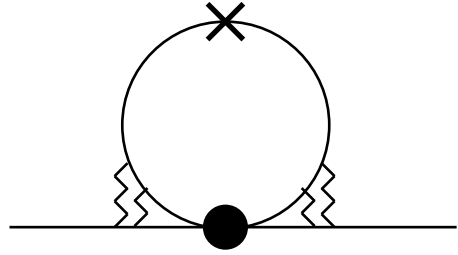

Fig. 1. Anomalous scaling enhancement of the *t* and *b* masses due to *X* boson exchange. Fourth family quarks are in the loop.

We now consider the following operators, some of which we label for our discussion of quark masses which follows.

$$(\bar{Q}_L D_R)\epsilon(\underline{\bar{Q}}_L U_R) \quad (\mathcal{C})$$
$$(\bar{Q}_L U_R)\epsilon(\underline{\bar{Q}}_L D_R) \quad (\tilde{\mathcal{C}})$$
$$(\underline{\bar{Q}}_L D_R)\epsilon(\underline{\bar{Q}}_L \underline{U}_R) \quad (\mathcal{D})$$
$$(\underline{\bar{Q}}_L U_R)\epsilon(\underline{\bar{Q}}_L \underline{D}_R) \quad (\tilde{\mathcal{D}}) \qquad (2)$$
$$(\bar{Q}_L \underline{D}_R)\epsilon(\bar{Q}_L U_R)$$
$$(\bar{Q}_L \underline{U}_R)\epsilon(\bar{Q}_L D_R)$$
$$(\underline{\bar{Q}}_L D_R)\epsilon(\bar{Q}_L \underline{U}_R)$$
$$(\underline{\bar{Q}}_L U_R)\epsilon(\bar{Q}_L \underline{D}_R) \qquad (3)$$

These operators cannot be generated on scales much higher than $\Lambda$ since they break $U(1)_V$. Assuming that the dynamics responsible for these operators is at scale $\Lambda$, they will show up as local operators in the effective theory below $\Lambda$. We will refer to them as $\Lambda$-operators, as opposed to the $\mathcal{B}$ and $\tilde{\mathcal{B}}$ operators which could be generated on some scale $\bar{\Lambda} \gg \Lambda$. We will refer to the $\mathcal{B}$ and $\tilde{\mathcal{B}}$ operators, and other such operators which respect $U(1)_V \times SU(2)_V$, as $\bar{\Lambda}$-operators. But it should be emphasized that because of anomalous scaling, the $\bar{\Lambda}$-operators when renormalized at scale $\Lambda$ are not necessarily suppressed relative to the $\Lambda$-operators.

---

[1] We note that other ways of attaching the *X* boson to the 4-fermion operator vertex would produce mixing with other operators, but at least at the one-loop level these other diagrams cancel among themselves.

All the $\Lambda$-operators in (2) and (3) are $SU(2)_V$ singlets; for example the $SU(2)_V$ indices on the $\mathcal{C}$ operator are contracted as follows, $(\bar{Q}_{Li}D_{Ri})\epsilon(\underline{\bar{Q}}_{Lj}U_{Rk}\epsilon_{jk})$. We can then expect that these operators are in an attractive channel with respect to $SU(2)_V$, and thus $SU(2)_V$ may dynamically generate these operators before $SU(2)_V$ breaks. At the same time, although the operators break $U(1)_V$ it appears that $U(1)_V$ does not resist the formation of the operators. At least the one-loop corrections to these operators involving the $U(1)_V$ boson cancel among themselves.

We now come to what may be the most important reason for why the various operators of the *LRLR* form are interesting: their structure implies nontrivial mass matrices. Let us illustrate this in the following submatrix of the $4 \times 4$ up-type matrix.

$$\begin{pmatrix} \underline{\bar{U}}_{L2}U_{R2} & \underline{\bar{U}}_{L2}\underline{U}_{R1} \\ \bar{U}_{L1}U_{R2} & \bar{U}_{L1}\underline{U}_{R1} \end{pmatrix} \tag{4}$$

We see that in the presence of the $\underline{\bar{D}}_{L1}D_{R1}$ mass (which is the $b'$ mass) that the operators labelled in (1) and (2) imply the following contributions.

$$\begin{pmatrix} 0 & \mathcal{D} \\ \mathcal{C} & \mathcal{B} \end{pmatrix} \tag{5}$$

The eigenvalues of this matrix are the $c$ and $t$ masses. The submatrix in the down sector describing the $s$ and $b$ masses is

$$\begin{pmatrix} 0 & \tilde{\mathcal{D}} \\ \tilde{\mathcal{C}} & \tilde{\mathcal{B}} \end{pmatrix} \tag{6}$$

The $\tilde{\mathcal{B}}$, $\tilde{\mathcal{C}}$, and $\tilde{\mathcal{D}}$ operators must be suppressed relative to the $\mathcal{B}$, $\mathcal{C}$, and $\mathcal{D}$ operators. This isospin breaking originates in the high energy physics, and it is not something we try to explain here.

The zero (1,1) entries in (5) and (6) could have received a contribution from the $SU(2)_V$ invariant operator

$$(\underline{\bar{Q}}_L D_R)\epsilon(\underline{\bar{Q}}_L U_R). \quad (\mathcal{E}) \tag{7}$$

But the formation of this operator is resisted by the $U(1)_V$ interaction, and thus may not arise dynamically in the same way as the other $\Lambda$-operators. Assuming this to be the case, a contribution to $\mathcal{E}$ still follows from the two-loop diagram in Fig. 2 involving the $\mathcal{B}$ $\bar{\Lambda}$-operator and $\Lambda$-operators in (2). By estimating the loop integrals in a manner we describe below we find that $\mathcal{E}$ is naturally suppressed relative to $\mathcal{B}$, $\mathcal{C}$ and $\mathcal{D}$. Note that the $\mathcal{E}$ operator is intrinsically isospin conserving and thus implies equal contributions to the (1,1) entries in (5) and (6), to the extent that the $t'$ and $b'$ masses are equal.

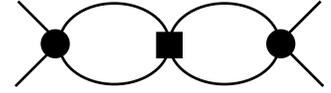

Fig. 2. The square (circle) i a $\bar{\Lambda}(\Lambda)$-operator.

As far as the resulting masses are concerned, we need the size of the operators when renormalized at a TeV. Strong $U(1)_X$ effects, and to a lesser extent QCD effects, will cause the various operators to scale differently as they are run down from $\Lambda$ to a TeV. Generally we expect that the most enhanced operators will be those with the most subscript 1 fields appearing in $U(1)_X$-invariant and Lorentz-scalar combinations. The result is that $\mathcal{B}$ is enhanced more than ($\mathcal{C}$, $\mathcal{D}$), while $\mathcal{E}$ is enhanced the least or not at all. A typical relative enhancement factor is $\Lambda/(1$ TeV). Putting all this together we find a natural hierarchical pattern of masses emerging.

Is this $\mathcal{E}$ entry, common to the two matrices, closer in size to the $c$ mass or the $s$ mass? The latter seems more natural; if $\mathcal{E}$ was of order the $c$ mass there would have to be some rather precise cancellation in the down-type mass matrix to produce a small enough $s$ mass. If $\mathcal{E}$ gives the $s$ mass then the $c$ mass must arise mostly through the mixing with the $t$, that is via the $\mathcal{C}$ and $\mathcal{D}$ entries. This then fixes the size of $\mathcal{C}$ and $\mathcal{D}$ entries to be of order $\sqrt{m_t m_c}$.

The appearance of this mixing may appear surprising, given that the third family feels a new gauge interaction which remains unbroken down to a TeV. The point is that the $\mathcal{C}$ and $\mathcal{D}$ operators break $U(1)_X$. The question then becomes, what contribution do the $\Lambda$-operators make to the $X$ mass? In more general terms, if a gauge symmetry is broken by a dynamical 4-point function rather than the usual dynamical 2-point function, what is the resulting gauge-boson mass? To deal with this question we must first describe the size of a $\Lambda$-operator. The coefficient of a $\Lambda$-operator near the $\Lambda$-scale may be written as some number over $\Lambda^2$, where $\Lambda$ is defined as the scale above which the $\Lambda$-operators are no longer local.[2] We will identify $\Lambda$ with the mass of gauge bosons in $(SU(2)_V \times U(1)_V)/U(1)_X$. Note that these massive gauge bosons must be involved in the nonperturbative generation of the $\Lambda$-operators, since only these fields couple to the subscript 2 field appearing in the $\Lambda$-operator. Other 4-fermion operators which are perturbatively generated have coefficients of order $g^2/\Lambda^2$, where $g^2 \approx 4\pi$ is typical for gauge interactions causing dynamical symmetry breaking. We will assume a similar result for the dynamically generated $\Lambda$-operators, and take the generic size of a $\Lambda$-operator to be $4\pi/\Lambda^2$. It cannot be much larger than this because of a unitarity upper bound of $\approx 8\pi/\Lambda^2$ on the size of a 4-fermion operator [7]. In addition, masses of the gauge bosons in $(SU(2)_V \times U(1)_V)/U(1)_X$ are related to the corresponding Goldstone boson decay constants, $\Lambda \approx gf$. This shows that the estimate $4\pi/\Lambda^2$ for the size of the $\Lambda$-operator is consistent with the naive dimensional analysis estimate, which would give $1/f^2$ [8].

---

[2] If it is a relevant operator on a lower scale $\Lambda'$ then its coefficient there is a number of order unity over $\Lambda'^2$.

Returning to the original question, how much do the $\Lambda$-operators contribute to a decay constant $f$? Roughly speaking this contribution is determined by a 3-loop diagram involving a $\Lambda$-operator 'vertex' and its hermitian conjugate, with the four fermion lines going from one vertex to the other. We are making an analogy with the derivation of the Pagels-Stokar [9] formula for $f_\pi$ which involves a single quark loop. By using $4\pi/\Lambda^2$ as the size of the $\Lambda$-operators, assigning a factor of $1/(4\pi)^2$ to each loop, and by cutting off the integrations at $\Lambda$ we find a contribution to $f$ of order $\Lambda/(4\pi)^2$. This is to be compared to the actual size of $f \approx \Lambda/\sqrt{4\pi}$. This means for example that our $\Lambda$-operators contribute very little to the $(SU(2)_V \times U(1)_V)/U(1)_X$ gauge boson masses. But since the $\Lambda$-operators also break $U(1)_X$ we may assume that they generate all or most of the $X$ mass, in which case

$$\frac{M_X}{g_X} \approx \frac{\Lambda}{(4\pi)^2} \ . \tag{8}$$

This then is our mechanism for the breaking of $U(1)_X$, via physics occurring at the higher scale $\Lambda$. We will assume in the following that $M_X/g_X$ is of order a TeV.

The basic assumption of our picture is that this breaking of $U(1)_X$ at a TeV somehow induces the $U(1)_X$-violating $t'$ and $b'$ masses. There is the dynamical question of why it is that the $t'$ and $b'$ masses correspond to $\bar{\underline{Q}}_{L1} Q_{R1}$ rather than the $U(1)_X$-conserving $\bar{Q}_{L1} Q_{R1}$ or $\bar{\underline{Q}}_{L1} \underline{Q}_{R1}$. The standard ladder approximation analysis would suggest that the latter is preferred. But even this analysis suggests that a critical coupling needs to be reached before any dynamical mass can form, and that this critical coupling is increased if there is only a finite range of momenta over which the attractive gauge interaction acts. In our case $X$ breaks at a TeV, and at a higher scale $\Lambda$ there may be new interactions which are repulsive in the $\bar{Q}_{L1}Q_{R1}$ or $\underline{\bar{Q}}_{L1}\underline{Q}_{R1}$ channels (as described in section 4). In addition if the required critical coupling is rather large, then the whole ladder approximation may break down and our naive intuitions about preferred channels may fail.

In summary we suppose that the $t'$ and $b'$ masses are generated by the isospin conserving $U(1)_X$ interactions, and that these masses are occurring in a channel distinct from the channels experiencing the attractive gauge interactions on higher energy scales.

## 3  Phenomenological Implications
### 3 a)  $\delta\rho$ and *Z-X* Mixing

We may now argue that the weak sector is shielded from the isospin breaking which is required in the operators responsible for the $t$ and $b$ masses [10]. In particular we must show that the $t'$–$b'$ mass splitting remains small, since the $t'$ and the $b'$ provide the dominant contribution to the $W$ and $Z$ masses. Although the various operators we have discussed reflect a badly broken isospin symmetry on high scales, none of them produce a direct contribution to $t'$–$b'$ mass-splitting. By a direct contribution we mean a diagram with only one insertion of the operator and with only the $t'$ and $b'$ masses involved. Operators reflecting the isospin breaking on high scales which would directly contribute to the splitting, for example $(\underline{\bar{Q}}_{L1}Q_{R1})\sigma_{3R}(\bar{Q}_{R1}\underline{Q}_{L1})$, are not expected to be enhanced by the $U(1)_X$ interaction since the factors $\underline{\bar{Q}}_{L1}Q_{R1}$ and $\bar{Q}_{R1}\underline{Q}_{L1}$ are not expected to have a large positive anomalous dimension. The situation would have been quite different if the $t'$ and $b'$ masses were instead identified with $\underline{\bar{Q}}_{L1}\underline{Q}_{R1}$.

The basic reason for why $\delta\rho$ may be sufficiently small is that the $t$ mass is still fairly small compared to the TeV mass expected for the $t'$ and $b'$. The $\mathcal{B}$ operator responsible for the $t$ mass is thus small in the sense that effects relying on multiple insertions of the operator are suppressed. In fact to produce a contribution to $\delta\rho$ there must be at least four insertions of the operator. For example a $t'$–$b'$ mass splitting would require two insertions, and $\delta\rho$ is proportional to the mass-splitting squared. The dangerous $(\bar{Q}_{R1}\gamma_\mu\sigma_3 Q_{R1})^2$ operator can cause a direct contribution to $\delta\rho$, but to produce this operator requires three loops and four insertions of the $\mathcal{B}$ operator. We avoid the large contributions to $(\bar{Q}_{R1}\gamma_\mu\sigma_3 Q_{R1})^2$ analogous to those arising from low-scale, isospin-violating ETC interactions, as we will explain more fully below.

The isospin breaking reflected in the $t$ mass implies what is perhaps the most interesting signatures associated with the $X$ boson, namely those which follow from the $Z$-$X$ mixing induced by a $t$ loop. As discussed previously [11], this causes the $Z$ couplings to the third family to be shifted. In light of recent data, this is interesting if the shift in the $Zb\bar{b}$ vertex roughly cancels the standard model correction to this vertex, which produces roughly a $-2\%$ correction. The latter correction also involves a $t$ loop along with an electroweak gauge boson. Thus for the two effects to roughly cancel we find three conditions. 1) The $X$ should have axial couplings to the $t$ to produce a mass mixing between the $Z$ and the $X$. 2) The $t$ and the $b$ should have the same sign axial $X$ coupling to produce an effect of the right sign. 3) $M_X/g_X$ should be similar to the mass-to-coupling ratio of the electroweak gauge bosons to produce an effect of comparable magnitude.

We have already seen how the model satisfies these conditions. The $t$ and $b$ quarks are to good approximation composed of the fields $Q_{L1}$ and $\underline{Q}_{R1}$, which have opposite $X$ charges, and thus the $t$ and the $b$ have equal axial couplings to the $X$. And we have argued that $M_X/g_X \approx 1$ TeV is reasonable, since we don't expect a large hierarchy between the $X$ mass and the $(t', b')$ masses. There is a lower bound on $M_X/g_X$ due to the fact that the $(t', b')$ masses do not respect $U(1)_X$; since the $t'$ and $b'$ masses can be related to the $W$ and $Z$ masses we can deduce that [11]

$$\frac{M_X^2}{g_X^2} \gtrsim \frac{2\sqrt{2}}{G} \ . \tag{9}$$

But it should be kept in mind that should $M_X/g_X$ be larger than 1 TeV, due to the poorly constrained contributions from the $\Lambda$-operators, then the effects we describe in this section can be severely reduced.

An effect similar to the $Z$–$X$ mixing has been noted in the context of more conventional ETC models [12]. Here a diagonal ETC gauge boson mixes with the $Z$ via a technifermion loop. Since the technifermion masses must respect isospin to good approximation, the ETC gauge boson must have isospin breaking couplings to the technifermions for this mixing to occur. It is not surprising then [13] that such a model leads to a large contribution to $\delta\rho$, via diagrams involving technifermion loops and the ETC gauge boson. In our case the *X* boson has isospin conserving couplings, which implies that the analogous diagrams involving the heavy, approximately degenerate, 4th-family quarks cause little problem. Instead the more important diagrams which contribute to $\delta\rho$ are those involving the *t*, shown in Fig. 3.

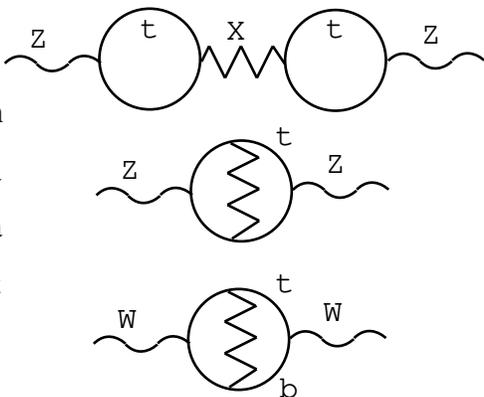

Fig. 3. Diagrams contributing to $\delta\rho$.

If the new physics contribution to the $Zb\bar{b}$ vertex is arranged to be the same in both cases, it may be seen that our contributions to $\delta\rho$ are suppressed relative to the ETC contributions by roughly the ratio of the *t* loop to the techniquark loop. We may write this ratio as $f_t^2/f^2$ where $f_t$ determines the *t* contribution to the *Z* mass and is defined along with *f* by

$$M_Z^2 = \frac{e^2}{4s^2c^2}(f^2 + f_t^2). \tag{10}$$

An NJL-type estimate of $f_t$ has been given elsewhere [14].

$$f_t^2 \approx \frac{3}{8\pi^2} m_t^2 \ln\left(\frac{(1\text{ TeV})^2}{m_t^2}\right) \tag{11}$$

We may also follow reference [14] to estimate our contribution to $\delta\rho$, where the three diagrams in Fig. 3 correspond to the three terms in parenthesis:

$$\delta\rho \approx (8 + 4 - 1)\left(\frac{g_X}{M_X}\right)^2 \frac{f_t^4}{f^2}$$

$$\approx 0.003 \text{ for } M_X/g_X = 1 \text{ TeV} \tag{12}$$

Our discussion in the next section will imply that the *X* has purely vector couplings to the $\tau$. This in fact is necessary to avoid upsetting the constraints on the *Z* partial width to the $\tau$ [11]. We thus conclude that the *X* boson couples to the following third family current,

$$J_\mu^X = \bar{t}(L_\mu - R_\mu)t + \bar{b}(L_\mu - R_\mu)b + \bar{\tau}(L_\mu + R_\mu)\tau + \bar{\nu}_\tau L_\mu \nu_\tau \tag{13}$$

where $R_\mu, L_\mu \equiv \gamma_\mu(1 \pm \gamma_5)/2$. The *Z* couplings to the third family are shifted by an amount $\delta g_Z Z^\mu J_\mu^X$ where

$$\delta g_Z = -r \frac{M_Z^2}{M_X^2} g_X. \tag{14}$$

*r* is the ratio of the *Z*–*X* mixing diagram involving a *t* loop, and the *Z* mass diagram involving *t'*

and $b'$ loops, and it is given by

$$r \approx \frac{g_X f_t^2}{(e/4cs)f^2} \quad . \tag{15}$$

The final result is

$$\delta g_Z \approx -\frac{4cs}{e} M_Z^2 \left(\frac{g_X}{M_X}\right)^2 \left(\frac{f_t}{f}\right)^2$$
$$\approx -0.003 \text{ for } M_X/g_X = 1 \text{ TeV} \tag{16}$$

The implications for electroweak corrections may be extracted from the previous work [11] in which we used $\delta g_Z = -0.0028$.[3] Of most interest is the shift in the $Zb\bar{b}$ vertex which would account for the present anomaly in this $Z$ partial width, as well as for the discrepency between the high and low energy values of $\alpha(M_Z)$ [11]. If these anomalies are real and are accounted for in this way, then the current $J_\mu^X$ in (13) implies that the $\tau$ asymmetry parameter, $\mathcal{A}_\tau$, is about 20% larger than $\mathcal{A}_e$ or $\mathcal{A}_\mu$. Such an effect is consistent with the forward-backward asymmetry data $\mathcal{A}_\tau/\mathcal{A}_{e,\mu} = 0.27 \pm 0.17$, but it is less consistent with the $\tau$ polarization data, $\mathcal{A}_\tau/\mathcal{A}_e = 0.02 \pm 0.09$ [15].

## 3 b) More anomalous gauge couplings

We will now describe a phenomenological consequence of operators of the form $(\bar{q}_L H_R)\epsilon(\bar{H}_L q_R)$. (In this subsection we return to the notation used in the introduction.) Closing off the heavy fermion lines and attaching a photon to that heavy fermion line yields a magnetic moment operator. Similarly attaching a gluon gives rise to a chromomagnetic moment operator

$$ig_s(\kappa/2m)\bar{q}\, T_a \sigma_{\mu\nu} k^\nu q\, G_a^\mu \quad . \tag{17}$$

We may compare the loop integral here to the one which gives the quark $q$ a mass $m$. We may assume that both integrals are dominated by similar momenta, since both integrals involve the momentum dependence of both the 4-fermion operator and the $H$ mass function. This leads to the estimate $\kappa \approx m^2/\lambda^2$ where $\lambda$ is of order a TeV. $\lambda$ is basically the 'scale of new physics'. In this section we shall be concerned with T-conserving physics, and ignore possible T-violating phases

---

[3] Our present estimate of $\delta g_Z$ involves a larger $M_X/g_X$ and a larger $r$ than our previous estimate, but the changes cancel.

which could be present in some of the $(\bar{q}_L H_R)\epsilon(\bar{H}_L q_R)$ operators.

[The situation is different in the walking technicolor context in which the mass generating operators of the form $(\bar{H}_L H_R)(\bar{q}_R q_L)$ receive anomalous scaling enhancement, but because of their spinor structure do not generate the (chromo)magnetic moment operator. The (chromo)magnetic moment operator may be generated by other ETC effects, but the effect will be smaller than in our case, basically by the amount by which masses are enhanced due to walking technicolor.]

Any additional momentum dependence of the chromomagnetic moment may be neglected since the $k^2$ of the gluon is typically much smaller than $\lambda^2$. Perhaps most interesting is the case of the top quark where the chromomagnetic moment will influence the production rate for ($gg$ or $q\bar{q}) \to t\bar{t}$. As shown in [16] the new contribution may start to be interesting for present experiments for $\lambda \lesssim 1$ TeV. Our 4-fermion operators play the same role as the exchange of colored scalars described in that reference and in [17].

The following flavor-changing couplings also exist,

$$ie/(2m_t)\bar{t}\,\sigma_{\mu\nu}k^\nu(C^{(\gamma,Z)} + D^{(\gamma,Z)}\gamma_5)c\,(A^\mu, Z^\mu) + h.c. \tag{18}$$

along with similar gluonic couplings. The couplings in (18) may provide a distinctive signature at $e^+e^-$ colliders [18]. If we define the ratio

$$R^{ct} \equiv \frac{\sigma(e^+e^- \to t\bar{c} + \bar{t}c)}{\sigma(e^+e^- \to \gamma \to \mu^+\mu^-)} \tag{19}$$

we find that as a function of the center of mass energy $\sqrt{s}$ that

$$R^{ct}(s) = \mathcal{K}^{ct}\frac{(s + 2m_t^2)(s - m_t^2)^2}{3s^2 m_t^2} \tag{20}$$

where we have neglected $m_c/m_t$ terms. The angular distribution is close to being proportional to $1 - \cos(\theta)^2$ where $\theta$ is the emission angle of the $t$ quark in the $e^+e^-$ center of mass frame. Because of an overall uncertainty in the couplings $C^\gamma$, $C^Z$, $D^\gamma$, $D^Z$ we do not bother to keep track of the individual contributions. In the section 2 we found that the $\mathcal{C}$ and $\mathcal{D}$ operators were a factor of $\sqrt{m_c/m_t}$ smaller than the $\mathcal{B}$ operator. Thus we might expect that a typical size of any one of these flavor-changing couplings is $\sqrt{m_c/m_t}\,(m_t^2/\lambda^2)$, and therefore $\mathcal{K}^{ct} \approx m_t^3 m_c/\lambda^4$.

Using this estimate, $R^{ct}(200 \text{ GeV})$ ranges from $5 \times 10^{-7}$ for $\lambda = 1$ TeV to $7 \times 10^{-6}$ for $\lambda = 0.5$ TeV. The same two numbers for $R^{ct}(500 \text{ GeV})$ are $2 \times 10^{-5}$ and $3 \times 10^{-4}$. Some of these numbers are easily at the detectable level if one can expect $10^6$–$10^7$ $\mu^+\mu^-$ events in a year of running. Even the limits obtainable from LEP200 would be interesting given the uncertainty in our estimates and in $\lambda$.

Similar flavor changing couplings have been considered [19] in the context of multi-Higgs models, in which loops involving the scalar fields can generate anomalous couplings. The removal of the gauge boson from such diagrams does **not** produce diagrams which produce the bulk of the fermion mass. In our picture they do (as in [17]), and thus there is a different relation between the size of the anomalous couplings and the fermion masses. Since masses in our picture are generated at one loop rather than from a Higgs vacuum expectation value, the implication is that the anomalous couplings are roughly a factor of $16\pi^2$ larger, when compared to the case when all scalar masses are of order $\lambda$.

For flavor-changing couplings to exist in our case it must not be possible diagonalize both the mass matrix and the 'matrix of couplings' simultaneously. We may argue that this is the case, because of the nontrivial momentum dependence of the $\mathcal{B}$, $\mathcal{C}$, and $\mathcal{D}$ operators for momenta of order a TeV. The point is that the $\mathcal{B}$ $\bar{\Lambda}$-operator and the ($\mathcal{C}$, $\mathcal{D}$) $\Lambda$-operators have quite different origins, and their different momentum behaviors is probed in different ways in the loops which generate the masses versus the loops which generate the anomalous couplings. Thus the mass matrix is not proportional to the matrix of couplings. We conclude that the flavor-changing $c$–$t$ coupling should exist with a size which is typically of order $\sqrt{m_c/m_t}$ times the anomalous $t$ coupling, as we assumed above. In the next section we will see that $\tilde{\mathcal{C}}$ and $\tilde{\mathcal{D}}$ entries may be smaller than $\sqrt{m_s/m_b}$ times the $\tilde{\mathcal{B}}$ entry, which leads to a suppression of the analogous flavor changing couplings in the down sector.

We should emphasize that among all the possible types of anomalous couplings, our picture is predicting that it is the (chromo)magnetic moments which dominate. This is because the structure of the operator $(\bar{q}_L H_R)\epsilon(\bar{H}_L q_R)$ is such that the resulting anomalous gauge couplings must change chirality. Thus anomalous couplings of the form $\bar{q}\gamma_\mu[A^{(\gamma,Z)}(q^2) + B^{(\gamma,Z)}(q^2)\gamma_5]q(A^\mu, Z^\mu)$ (and similarly for the gluon $\mathcal{G}$, with $A^\gamma(0) = B^\gamma(0) = A^\mathcal{G}(0) = B^\mathcal{G}(0) = 0$) are not generated. These latter couplings are generated in multi-Higgs models, and we note that the presence of these couplings would affect the $s$ dependence in (20). There are also many other 4-fermion operators generated in our model with a variety of effects, but they are suppressed compared to the *t*-mass generating operator since, as we shall see, they do not enjoy the same anomalous scaling enhancement.

## 4 The full 4-family model
## 4 a) Leptons and more $\bar{\Lambda}$-operators

We now consider the theory further at scales $\Lambda$ and above, where the gauge symmetry is

$$U(1)_V \times SU(2)_V \times SU(3)_C \times SU(2)_L \times U(1)_Y . \tag{21}$$

The complete fermion content consists of the two sets of fermions $(Q, L)$ and $(\underline{Q}, \underline{L})$. Each set is a standard family of quarks and leptons, where each such 'quark' $Q$ or $\underline{Q}$ and 'lepton' $L$ or $\underline{L}$ carries an $SU(2)_V$ index. $(Q, L)$ transform as $(+, 2)$ under $U(1)_V \times SU(2)_V$ while $(\underline{Q}, \underline{L})$ transform as $(-, \bar{2})$. A right-handed neutrino Majorana mass $N_R \underline{N}_R$ is invariant under the above gauge symmetry. We therefore expect that right-handed neutrinos do not exist in the theory at scale $\Lambda$, since it is natural for them to have a much larger mass.

Another anomaly free $U(1)$ may also be gauged above $\Lambda$; we refer to a $U(1)_A$ under which the $(Q, L)$ and $(\underline{Q}, \underline{L})$ have equal and opposite **axial** charges. This interaction serves the useful purpose of resisting the formation of the $U(1)_V \times SU(2)_V$ singlet masses $\bar{Q}_L Q_R$ and $\underline{\bar{Q}}_L \underline{Q}_R$. The mass of the $U(1)_A$ boson could be of order $\Lambda$ or larger. We note that its effects are isospin symmetric and that the various $\bar{\Lambda}$-operators we have discussed are neutral under $U(1)_A$. We will leave the existence of the $U(1)_A$ as an open question for now.

The gauge group and fermion content we have described would have a number of global symmetries leading to unwanted Goldstone bosons. As mentioned in the introduction, this is not a problem in the presence of an appropiate set of relevant $\bar{\Lambda}$-operators. It turns out that the following two-quark-two-lepton $\bar{\Lambda}$-operators are sufficient to break all the global symmetries.

$$(\underline{\bar{L}}_L \underline{E}_R)\epsilon(\bar{Q}_L U_R) \quad (\mathcal{F})$$
$$(\bar{L}_L E_R)\epsilon(\underline{\bar{Q}}_L \underline{U}_R) \tag{22}$$
$$(\bar{L}_L E_R)\epsilon(\bar{Q}_L U_R)$$
$$(\underline{\bar{L}}_L \underline{E}_R)\epsilon(\underline{\bar{Q}}_L \underline{U}_R) \quad (\mathcal{G}) \tag{23}$$

Their $U(1)_V \times SU(2)_V$ structure is similar to the $\mathcal{B}$ and $\tilde{\mathcal{B}}$ operators in (1), and thus we may expect similar anomalous scaling enhancement.[4] There are no purely leptonic operators of the *LRLR* form because of the absence of the right-handed neutrinos.

These operators will play another useful role, by providing some crosstalk between the quark and lepton mass matrices. In particular we see that the $\tau'$ mass can feed into the quark mass matrix if the $\tau'$ mass is of the form $\bar{E}_{L1} E_{R1}$ or $\underline{\bar{E}}_{L1} \underline{E}_{R1}$, which is a form different than that of the heavy quark masses. We note that the absence of color interactions will modify the influence of various 4-fermion operators in the lepton sector. Because of this different dynamics, we shall feel free to assume that the $\tau'$ mass corresponds to a dynamical $\underline{\bar{E}}_{L1} \underline{E}_{R1}$ mass. The implication is that the $\tau$ is composed of the $E_{L1}$ and $E_{R1}$ fields, which leads to the purely vector X coupling to the $\tau$ we noted in the previous section.

## 4 b)  Mass Matrices

We now consider the full $4 \times 4$ mass matrices with the following elements, where we have made explicit the $SU(2)_V$ indices.

$$\begin{vmatrix} \bar{Q}_{L2}\underline{Q}_{R2} & \bar{Q}_{L2}Q_{R2} & \bar{Q}_{L2}\underline{Q}_{R1} & \bar{Q}_{L2}Q_{R1} \\ \underline{\bar{Q}}_{L2}\underline{Q}_{R2} & \underline{\bar{Q}}_{L2}Q_{R2} & \underline{\bar{Q}}_{L2}\underline{Q}_{R1} & \underline{\bar{Q}}_{L2}Q_{R1} \\ \bar{Q}_{L1}\underline{Q}_{R2} & \bar{Q}_{L1}Q_{R2} & \bar{Q}_{L1}\underline{Q}_{R1} & \bar{Q}_{L1}Q_{R1} \\ \underline{\bar{Q}}_{L1}\underline{Q}_{R2} & \underline{\bar{Q}}_{L1}Q_{R2} & \underline{\bar{Q}}_{L1}\underline{Q}_{R1} & \underline{\bar{Q}}_{L1}Q_{R1} \end{vmatrix} \tag{24}$$

With the operators we have discussed the up-type mass matrix takes the following form.

$$\begin{vmatrix} 0 & \mathcal{F}_2 & 0 & 0 \\ \mathcal{G}_2 & \mathcal{E} & \mathcal{D} & 0 \\ 0 & \mathcal{C} & \mathcal{B} & \mathcal{F}_1 \\ 0 & 0 & \mathcal{G}_1 & \mathcal{A} \end{vmatrix} \tag{25}$$

The entry $\mathcal{A}$ comes directly from the dynamical $\underline{\bar{Q}}_{L1} Q_{R1}$ mass. The $\mathcal{F}$ and $\mathcal{G}$ operators, labelled in (22) and (23), are feeding mass into the quark sector from the $\tau'$ mass (subscripts on the $\mathcal{F}$ and $\mathcal{G}$ refer to the $SU(2)_V$ indices appearing in the corresponding operator). We expect $\mathcal{B}$ to be enhanced relative to ($\mathcal{F}$, $\mathcal{G}$), both because of QCD renormalization effects and because the $b'$ mass is likely larger than the $\tau'$ mass. And ($\mathcal{F}_1$, $\mathcal{G}_1$) will be enhanced relative to ($\mathcal{F}_2$, $\mathcal{G}_2$) due to strong $U(1)_X$ scaling effects below the scale $\Lambda$. Note that we have a nonsymmetric mass matrix since, although we expect that $\mathcal{C} \approx \mathcal{D}$, there is no reason to expect that $\mathcal{F} \approx \mathcal{G}$.

---

[4] The operators in (22) are $U(1)_A$ symmetric while those in (23) are not. We will also find below that the former should be larger than the latter.

The down-type matrix takes the form

$$\begin{vmatrix} \mathcal{H} & 0 & I & 0 \\ 0 & \mathcal{E} & \tilde{\mathcal{D}} & 0 \\ I & \tilde{\mathcal{C}} & \tilde{\mathcal{B}} & 0 \\ 0 & 0 & 0 & \mathcal{A} \end{vmatrix} \qquad (26)$$

where the new entries correspond to

$$(\bar{Q}_L \underline{U}_R)\epsilon(\bar{Q}_L \underline{D}_R) \qquad (\mathcal{H})$$
$$(\bar{\underline{E}}_{R1} \underline{E}_L)\epsilon(\bar{D}_L \underline{D}_R) \qquad (I) \qquad (27)$$

The $\mathcal{H}$ operator is feeding mass from $t$ to $d$. It arises in a way very similar to the $\mathcal{E}$ operator, except that the two-loop diagram involves the $\Lambda$-operators in (3) rather than (2). It also has a partner which would feed mass from $b$ to $u$, but since $m_b \ll m_t$ we have neglected this effect in the up-type matrix. Thus we have a situation in which the $d$ mass is connected with the $t$ mass, while the $u$ mass is connected to the $\tau'$ mass. Noting that the $\mathcal{E}$ operator is feeding mass from $t'$ to $s$, and the fact that the $\mathcal{H}$ and $\mathcal{E}$ operators are expected to be of similar size, leads to the not completely ridiculous relation

$$m_d/m_t \approx m_s/m_{t'}. \qquad (28)$$

The $I$ operator is feeding mass down from the $\tau'$. This operator arises from a loop, as shown in Fig. 4, involving the first $\Lambda$-operator in (3) and the $\mathcal{F}$ operator. The $\tilde{\mathcal{C}}$ and $\tilde{\mathcal{D}}$ operators may be generated by a similar loop, if they are not dynamically generated directly. In that case the two operators in the loop would be a $\mathcal{C}$ or $\mathcal{D}$ operator and a possible $\bar{\Lambda}$-operator $(\bar{Q}_L Q_R)(\bar{\underline{Q}}_R \underline{Q}_L)$. In constructing loop diagrams we note that the right-handed neutrinos are not in the theory at the $\Lambda$ scale and thus we may ignore diagrams involving internal right-handed neutrinos and $\Lambda$-operators.

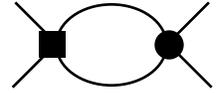

Fig. 4.

Of particular interest for the suppression of flavor changing neutral currents is the possible suppression of the (1,2) and (2,1) entries in the down-type mass matrix. There are contributions from the $\tau'$ through a loop involving two $\bar{\Lambda}$-operators where one of the $\bar{\Lambda}$-operators is either $\tilde{\mathcal{B}}$ or $\mathcal{G}$. But the latter operators are suppressed; we have seen that $\tilde{\mathcal{B}}$ is smaller than $\mathcal{B}$ and we will see that $\mathcal{G}$ is smaller than $\mathcal{F}$. Another possible contribution involves a loop with two $\Lambda$-operators where one operator is different from any of those discussed above, either $(\underline{\bar{L}_I}E_R)(\bar{Q}_L\underline{U}_R)$ or $(\underline{\bar{L}_I}E_R)(\bar{Q}_L\underline{U}_R)$. We will assume for now that all these contributions to the (1,2) and (2,1) entries are small, which means that Cabbibo mixing mostly originates in the up sector. Finally the (1,4) and (4,1) entries can be generated by the $\Lambda$-operators in (3); but the position of these entries makes them quite unimportant and we set them to zero.

## 4 c) Mass Mixings

Do the mass matrices of the form presented allow for a realistic quark mass spectrum and Kobayashi-Maskawa (KM) mixing matrix? We will find that they do and that the elements of the mass matrices are quite constrained. We can illustrate this with some sample matrices, but we are of course not claiming to show that the model actually does produce these matrices. We choose the up-type matrix to be the following, which gives the following set of masses when diagonalized: (0.022, 0.74, 160, 1000) GeV.[5]

$$M^u = \begin{vmatrix} 0 & .16 & 0 & 0 \\ -.01 & .1 & 10 & 0 \\ 0 & -10 & 160 & 16 \\ 0 & 0 & -1 & 1000 \end{vmatrix} \quad (29)$$

The (3,3) entry is fixed by the $t$ mass. The (2,2) entry is the same as in the down-type matrix and is thus basically fixed by the $s$ mass. Then the (2,3) entry is fixed in order to obtain the $c$ mass through mixing with the $t$. The (1,2) entry is determined if we want the bulk of Cabibbo mixing to occur in the up sector, and finally the (2,1) entry is determined by the $u$ mass. Here we see how small $\mathcal{G}_2$ is compared to $\mathcal{F}_2$. The (3,4) and (4,3) entries coming from $\mathcal{F}_1$ and $\mathcal{G}_1$ are of little consequence, and we have set them equal to 100 times the $\mathcal{F}_2$ and $\mathcal{G}_2$ entries respectively. The resulting left and right-handed mixing matrices are

$$L^u = \begin{vmatrix} .98 & .22 & 0 & 0 \\ -.22 & .97 & .062 & 0 \\ .013 & -.061 & 1.0 & .016 \\ -.0002 & .0010 & -.016 & 1.0 \end{vmatrix} \tag{30}$$

$$R^u = \begin{vmatrix} 1.0 & -.013 & 0 & 0 \\ .013 & 1.0 & -.062 & -.0002 \\ .0008 & .062 & 1.0 & .0016 \\ 0 & 0 & -.0016 & 1.0 \end{vmatrix} \tag{31}$$

---

[5] These are reasonable masses since they are renormalized at a TeV.

For the down-type matrix we take

$$M^d = \begin{vmatrix} .005 & 0 & -.015 & 0 \\ 0 & .1 & .07 & 0 \\ .015 & -.07 & 3 & 0 \\ 0 & 0 & 0 & 1000 \end{vmatrix} \tag{32}$$

Here the physical masses are very similar to the diagonal components of this matrix. The (2,3) and (3,2) entries are determined such that we obtain the correct $V_{cb}$, while the (1,3) and (3,1) entries are determined by $V_{ub}$. For the left-handed mixing matrix we have

$$L^d = \begin{vmatrix} 1.0 & -.0035 & -.0050 & 0 \\ .0036 & 1.0 & .023 & 0 \\ .0049 & -.023 & 1.0 & 0 \\ 0 & 0 & 0 & 1.0 \end{vmatrix} \tag{33}$$

$R^d$ is the same except that the off-diagonal entries have opposite sign. The main interest here is that the mixing between the $d$ and $s$ quarks is small, which provides a useful suppression of $K^0$–$\bar{K}^0$ mixing induced by the $\Lambda$-scale physics. We have thus managed to produce a realistic KM mixing matrix, $V_{KM} \equiv L^{u\mathrm{T}} L^d$.

$$V_{KM} = \begin{vmatrix} .98 & -.22 & .0030 & -.0002 \\ .22 & .97 & -.040 & .0010 \\ .0051 & .039 & 1.0 & -.016 \\ 0 & -.0004 & .016 & 1.0 \end{vmatrix} \tag{34}$$

Our conclusion is that if the 4-fermion operators we considered are actually the dominant ones, then the quark mixing matrices are fairly well determined.

For completeness we briefly consider lepton masses. We have noted that the $\underline{\bar{E}}_{L1}\underline{E}_{R1}$ mass ($\tau'$ mass) can play an important role in feeding mass into the quark sector. There are three operators which can feed the $\tau'$ mass down to the other three charged leptons.

$$(\underline{\bar{E}}_{R1}\underline{E}_{L1})(\bar{E}_{L1}E_{R1}) \quad (\underline{\bar{E}}_{R1}\underline{E}_{L1})(\bar{E}_{L2}E_{R2}) \quad (\underline{\bar{E}}_{R1}\underline{E}_{L1})(\underline{\bar{E}}_{L2}\underline{E}_{R2}) \tag{35}$$

The existence of the first two operators is implied through one-loop effects involving the $\bar{\Lambda}$-operators in (22) and (23). The first is enhanced through $U(1)_X$-induced anomalous scaling relative to the other two. The third operator can actually be generated by an explicit gauge interaction, namely a broken $SU(2)_V$ generator. But effects very similar to those generating the second operator can also contribute to the third, and so it appears that the third operator must be larger than the second. Thus we speculate that the three operators listed are generating the $\tau$, $e$, and $\mu$ masses respectively. We note that unlike the quark sector, the operators in (35) are not of the *LRLR* form, and therefore do not directly contribute to anomalous magnetic moments for $e$, $\mu$, and $\tau$. As for the neutrinos, a Majorana $\underline{N}_{L1}\underline{N}_{L1}$ mass would be a $\nu_{\tau'}$ mass. If this was somehow dynamically generated then we would be left with three light neutrinos $(N_{L1}, \underline{N}_{L2}, N_{L2}) \equiv (\nu_\tau, \nu_\mu, \nu_e)$ which do not receive mass via any 4-fermion operator.

## 5 Conclusion

This work originated in a proposal [10] to have anomalous scaling enhance the 4-fermion operator responsible for the *t* mass more than other dangerous isospin-violating operators. The new strong gauge interaction, the $U(1)_X$, responsible for the anomalous scaling must couple to the third family and it must be broken close to a TeV. Electroweak symmetry breaking is induced by a TeV mass for fourth family quarks, where this mass must occur in a channel which does not respect $U(1)_X$. We have argued that the physics responsible for breaking $U(1)_X$ can lie at a scale $\Lambda$ higher than a TeV, when the symmetry breaking order parameter involves 4-point functions rather than the usual 2-point functions.

At scale $\Lambda$ the $U(1)_X$ becomes embedded in a larger gauge gauge symmetry, a $U(2)$, which involves the two light families as well. The full quark mass matrix can then be related to various 4-fermion operators present at the $\Lambda$-scale. Hierarchies develop in the mass matrices for three reasons. 1) The various operators have different numbers of fields to which the $U(1)_X$ couples, and thus are enhanced by varying amounts due to anomalous scaling induced by the $U(1)_X$. 2) Some entries in the mass matrices receive mass fed down from a $t$ or $\tau'$ rather than the heavier $t'$ or $b'$. 3) Some entries only receive contributions from loops involving more than one 4-fermion operator. All of this relies on the existence of chirality-changing operators of the *LRLR* form. These couplings in turn lead to anomalous couplings of the photon, *Z*, and gluon to quarks, with perhaps the most interesting signatures involving the production of $t\bar{t}$ and $(t\bar{c} + \bar{t}c)$. The *Z* will also mix with the *X*, which can cause interesting shifts in the *Z* coupling to the third family. We note that it is difficult to produce the *X* boson directly in colliders, because its tree level couplings to light fermions are only those induced by the small mass mixing between the light and heavy families.

This work leaves a number of issues unresolved. Perhaps the main question has to do with the ultimate origin of the $\bar{\Lambda}$-operators, especially the $\mathcal{B}$ and $\tilde{\mathcal{B}}$ operators which are responsible for the *t* and *b* masses. Their chirality breaking structure indicates that they have a dynamical origin and, as suggested in the introduction, the physics responsible could lie at some very high mass scale. This same dynamics would also be responsible for the isospin breaking reflected in the coefficients of these operators. The fact that isospin breaking can be generated dynamically in this way can be demonstrated by the study of a toy scalar field model. Namely we may consider a $SU(2)_L \times SU(2)_R$ symmetric scalar field potential involving scalar fields having the same $SU(2)_L \times SU(2)_R$ indices as the $\mathcal{B}$ and $\tilde{\mathcal{B}}$ operators, and then show for a range of parameters that isospin is broken via the breaking of $SU(2)_R$ in the desired manner. This exercise was carried out in the appendix in [20].

We should also stress that we have made the minimal choice for the new gauge interactions, which exhibit the following symmetry breaking pattern, $U(2)_V \to U(1)_X \to$ nothing. These gauge symmetries could be enlarged for example to the symmetry breaking pattern $U(4) \to SU(3) \to SU(2)$,[6] and most of the discussion of this paper would still carry through. The main difference would be that a new sector of fermions with unbroken $SU(2)$ gauge interactions would remain which, unlike a technicolor sector, need not play any substantial role in electroweak symmetry

---

[6] This is the case proposed in our earlier work (for example see [19] ).

breaking.

Finally, we bring the readers attention to some recent related work. Of interest for our discussion in the introduction is the study [21] of the nontrivial infrared fixed points in nonabelian gauge theories. In particular the number of flavors with respect to the $SU(4)$ mentioned in the last paragraph is 15 (two families without right-handed neutrinos), which puts the infrared fixed point suggested in [21] very close to the critical coupling needed for chiral symmetry breaking. In other work [22] we have found that the dynamical right-handed fourth-family Majorana neutrino mass may make a useful negative contribution to the electroweak correction parameter $\delta\rho$.

## Acknowledgements

I thank T. Kugo and the Yukawa Institute for Theoretical Physics, where some of this work was completed, for their hospitality and support. I thank M. Luke, V. Miransky and M. Ramana for useful discussions. This research was also supported in part by the Natural Sciences and Engineering Research Council of Canada.

## References

[1]  W. Bardeen, C.N. Leung, S.T. Love, Nucl. Phys. **B273** (1986) 649.
[2]  A. Cohen and H. Georgi, Nucl. Phys. **B314** (1989) 7.
[3]  B. Holdom, Phys. Rev. Lett. **62** (1989) 338;  U. Mahanta, Phys. Lett. **B225** (1989) 181. Footnote 3 in the last reference should be ignored.
[4]  R. S. Chivukula, Phys. Rev. Lett. **61** (1988) 2657.
[5]  B. Holdom, Phys. Rev **D24** (1981) 1441; B. Holdom, Phys. Lett. **B150** (1985) 301; K. Yamawaki, M. Bando, and K. Matumoto, Phys. Rev. Lett. **56** (1986) 1335; T. Appelquist, D. Karabali, and L.C.R. Wijewardhana, Phys. Rev.Lett. **57** (1986) 957; T. Appelquist and L.C.R. Wijewardhana, Phys. Rev **D35** (1987) 774; T. Appelquist and L.C.R. Wijewardhana, Phys. Rev **D36** (1987) 568.
[6]  B. Holdom and G. Triantaphyllou, Phys. Rev. **D51** (1995) 7124; B. Holdom and G. Triantaphyllou,  to be published in Phys. Rev. D, hep-ph/9505364; P. Maris and Qing Wang, hep-ph/9511403.
[7]  T. Asaka , N. Maekawa , T. Moroi , Y. Shobuda , and Y. Sumino,  talk given at 5th Workshop on Japan Linear Collider, Tsukuba, Japan, Feb. 1995, hep-ph/9505371.
[8]  A. Manohar and H. Georgi, Nucl. Phys. **B234** (1984) 189; H. Georgi and L. Randall, Nucl. Phys. **B276** (1986) 241.
[9]  H. Pagels and S. Stokar, Phys. Rev. **D20** (1979) 2947.

[10]  B. Holdom, Phys. Lett. **B336** (1994) 85.
[11]  B. Holdom, Phys. Lett. **B339** (1994) 114; **B351** (1995) 279.
[12]  G.-H. Wu, Phys. Rev. Lett. **74** (1995) 4137; C.-X. Yue, Y.-P. Kuang, G.-R. Lu, L.-D. Wan, Phys.Rev. **D52** (1995) 5314; K. Hagiwara, N. Kitazawa, Phys. Rev. **D52** (1995) 5374.
[13]  T. Yoshikawa, (Hiroshima U.) HUPD-9514, Jun 1995, hep-ph/9506411 .
[14]  R. S. Chivukula, B.A. Dobrescu, J. Terning, Phys.Lett. **B353** (1995) 289.
[15]  LEP results presented at the Europhysics Conference in Brussels, July 1995, and the Lepton-Photon Conference in Beijing, August 1995.
[16]  D. Atwood, A. Kagan, T.G. Rizzo, SLAC-PUB-6580 (1994) hep-ph/9407408.
[17]  A. Kagan, Phys. Rev. **D51** (1995) 6196.
[18]  D. Atwood, L. Reina, A. Soni , SLAC-PUB-95-6927 (1995) hep-ph/9506243.
[19]  M. Luke and M. J. Savage, Phys. Lett. **B307** (1993) 387.
[20]  B. Holdom, talk presented at the Yukawa International Seminar '95, August 1995, Kyoto, hep-ph/9510249.
[21]  T. Appelquist, J. Terning, L.C.R. Wijewardhana, Yale Preprint YCTP-P2-96, hep-ph/9602385.
[22]  B. Holdom, Toronto preprint UTPT-96-01, hep-ph/9602248.